# Gravitational Microlensing by Random Motion of Stars: Analysis of Light Curves

Joachim Wambsganss

Max-Planck-Institut für Astrophysik, Karl-Schwarzschildstr. 1, 85740 Garching, Germany

and

Astrophysikalisches Institut Potsdam An der Sternwarte 16, 14482 Potsdam, Germany (current address)

e-mail: jwambsganss@aip.de

and

Tomislav Kundić

Princeton University Observatory, Peyton Hall, Princeton, NJ 08544 USA

e-mail: tomislav@astro.princeton.edu

## ABSTRACT

We present a quantitative analysis of the effect of microlensing caused by random motion of individual stars in the galaxy which is lensing a background quasar. We calculate a large number of magnification patterns for positions of the stars slightly offset from one frame to the next, and in this way obtain light curves for fixed quasar and galaxy positions, only due to the change in the relative star positions. These light curves are analyzed to identify microlensing events, which are then classified with respect to height, duration, and slope. These random motion microlensing events are compared with the corresponding ones caused by the bulk motion of the galaxy.

We find that microlensing events produced by random motion of stars are shorter, steeper, and more frequent than bulk motion events, assuming the velocity dispersion of the stars equals the bulk velocity of the galaxy. The reason for this difference is that in the case of random motion caustics can move with an arbitrarily high velocity, producing very short events, whereas in the comparison case for bulk motion a microlensing event can never be shorter than it takes a fold caustic, which moves with the velocity of the lensing galaxy projected onto the quasar plane, to cross the quasar. An accompanying video illustrates these results. For three different values of the surface mass density $\kappa$, it shows time sequences of 1000 magnification patterns for slowly changing lens positions, together with the positions and velocity vectors of the microlensing stars. A short version of the video is available as MPEG movie under anonymous ftp at astro.princeton.edu, in the directory `jkw/microlensing/moving_stars`.

# 1. Introduction

Gravitational microlensing, the lensing effect by objects of roughly stellar masses on background sources, has recently been reported in three different regimes: in searches for compact halo objects, in quasar variability monitoring programs and in studies of multiply imaged quasars.

All three groups monitoring stars in the Magellanic Clouds or in the bulge of the Galaxy, i.e. the EROS group (Aubourg et al. 1993), the MACHO collaboration (Alcock et al. 1993), and the OGLE team (Udalski et al. 1993, 1994) reported the detection of microlensing events. At this point it is not yet clear whether the frequency of events is compatible with all the dark mass in the halo in form of compact objects, but it is encouraging that one can detect microlensing in the limit of very small optical depth (of order $10^{-6}$), when only one star in a million is affected by microlensing at any time.

Hawkins (1993) reports that over a period of almost two decades basically all quasars in his sample are variable. From the characteristics of the variability he interprets this as a result of microlensing by cosmologically distributed sub–solar mass objects along the lines of sight to the quasars. If the lensing interpretation turns out to be correct and the variability of quasars in the sample is due to microlensing, then the optical depth of lensing objects with masses $M \leq 10^{-3} M_\odot$ is of order unity; higher masses are excluded (Schneider 1993, Dalcanton et al. 1993)

The optical depth is also of order unity in multiply imaged quasars lensed by foreground galaxies. Two or more macroimages of background quasars are produced by the potential of the galactic core, but each individual image is further microlensed by ordinary stars inside the lensing galaxy. The best known example of this phenomenon is the quadruple quasar 2237+0305 (Huchra et al. 1985), where at least five microlensing events have been detected so far (Irwin et al. 1989, Corrigan et al. 1991, Pen et al. 1993, and Houde & Racine 1994). There are indications that microlensing has been observed in three other multiple lens systems as well (cf. Wambsganss 1993): in the double quasar 0957+561 (Schild & Smith 1991, Schild & Thomson 1993), in the "clover leaf" H1413+117 (Arnould et al. 1993), and in PG 1115+080 (Angonin et al. 1993, Schechter 1993).

In this paper we limit ourselves to microlensing by a single plane of deflectors at optical depth near unity, which can be applied most directly to multiply lensed quasars. In theoretical studies of microlensing one typically calculates light curves for a random, but *static* distribution of stars. The two standard methods are "ray-shooting" (Wambsganss 1990) and the inversion of the lens equation (Witt 1993, Lewis et al. 1993). Aside from the original paper by Paczyński (1986) almost all subsequent studies of microlensing considered fixed star positions. Only recently the effect of microlensing due to motion of individual stars was studied in somewhat more detail (Kundić & Wambsganss 1993, Kundić et al. 1993, Schramm et al. 1993).

Here we present a further investigation of the effect of random motion of stars on the statistics of microlensing variability. In particular, we analyze light curves of background sources microlensed by a field of stars whose relative positions are changing with time due to random motion, and compare them with the light curves produced by a stellar field in bulk motion. Our motivation is the fact that in microlensing calculations where a lightcurve was determined by relative motion between the quasar, the galaxy (with fixed relative star positions) and the observer, there could be no feature in the lightcurve shorter than the time it takes a caustic to cross the quasar. This has important consequences on the size of the quasar (Rauch and Blandford 1991, Jaroszyński et al. 1992). The velocity of the caustic is basically the transverse velocity of the lensing galaxy, projected onto the quasar plane. For random motion of lensing stars, however, the caustics can move with velocities much higher than the projected velocity of an individual star. This means that one can be quantitatively wrong if one determines the size of the quasar continuum region from an assumed transverse velocity of the lensing galaxy. Our goal is to study this effect quantitatively and to assess the frequency and the importance of "high-speed" caustics.

In Section 2 we describe our method, in Section 3 we present light curves and distributions of durations of peaks and slopes of light curves. Finally in Section 4 we discuss the consequences of these findings for microlensing observations and determinations of the sizes of quasars. There is an accompanying video presentation illustrating our results; it shows the changes in the magnification patterns due to random motion of lenses, the positions and velocities of the respective lenses, and sample light curves for microlensing by random motion of stars and by bulk motion of the lensing galaxy. The video that comes with this paper is described in the Appendix, as is the abbreviated MPEG version that can be obtained via anonymous ftp.

## 2. Method

This paper is a continuation of the work presented in Kundić & Wambsganss (1993). The method we use is the same as in the first paper: the backwards ray-shooting (Wambsganss 1990). We follow light rays deflected by stars in the lensing galaxy to the source plane and collect them in an array of 500×500 pixels. The density of light rays at every point in the source plane is directly proportional to the magnification at this location (therefore this pixel field is called magnification pattern). The sidelength of the magnification pattern is $20\xi_0$, where the Einstein radius $\xi_0$ is given by

$$\xi_0 = \sqrt{\frac{4GM}{c^2}\frac{D_{\rm d}D_{\rm ds}}{D_{\rm s}}} \quad . \tag{1}$$

Here $G$ is the gravitational constant, $c$ is the velocity of light, $D_d, D_s, D_{ds}$ are the angular diameter distances observer-lens, observer-source, lens-source, and $M$ is the mass of the lens. In this paper, all lenses have equal mass $M$.

We assign a two-dimensional transverse velocity $\boldsymbol{v}_i$ to each lens, random in direction and with an absolute value drawn from a two-dimensional Maxwellian distribution, which has the velocity dispersion equal to the bulk motion velocity of the galaxy. After a magnification pattern is obtained, we take a small time step $\Delta t$, move each lensing star by an amount corresponding to $\Delta \boldsymbol{x}_i = \boldsymbol{v}_i \Delta t$, and calculate a new magnification pattern for this slightly different lens configuration. This calculation is repeated for a large number of small time steps to obtain a quasi–continuous change in the magnification pattern. The lens field we use is much larger than the size necessary to cover the magnification pattern: the fact that some stars move out of the field during the sequence is statistically compensated by other stars moving into the field at other locations and from other directions.

In this way, 500 different magnification patterns are computed. The timestep $\Delta t$ is adjusted in such a way that the average displacement of a star is one pixel per timestep, or $\langle |\Delta \boldsymbol{x}_i| \rangle = 0.04$ $\xi_0$. In the case of Huchra's lens 2237+0305 (cf. Kundić & Wambsganss 1993), lenses have masses of M = 0.1 M$_\odot$ and a velocity dispersion of 215 km/sec (Foltz et al. 1992), so that a timestep corresponds to about two months. Whereas a standard lightcurve for bulk motion is usually obtained by a one-dimensional cut through a magnification pattern, a stack of frames allows us to determine light curves for random motion of stars as well: Recording the magnification at a fixed pixel in the receiving field for the time sequence of magnification patterns corresponds to a fixed relative position between the quasar, the galaxy and the observer, in which only the stars inside the lensing galaxy move. The fact that we consider random motion rather than some organized velocity is justified, because only the velocity component projected onto the sky is relevant for microlensing. Even in a spiral galaxy (except when it is seen face on), stellar velocities are well mixed in projection.

This "stack" of magnification patterns is illustrated in Figure 1. Each square represents a magnification pattern for slightly different lens positions. The arrowed lines labelled "bulk" and "random" mark light curves obtained with the two methods: bulk motion (the horizontal line along the first magnification pattern) and random motion (dotted line crossing the stack).

In fact, we can obtain a two–dimensional array that is a collection of random motion light curves, by cutting out the same pixel row from each magnification pattern, as illustrated by the shaded square. Such a time-magnification pattern is shown in Figure 2a: here a lightcurve for pure random motion is a horizontal cut, whereas a lightcurve for standard bulk motion is a vertical cut. A cut at a certain angle $\phi$ (relative to the horizontal line) corresponds to a combination of bulk motion and random motion with the ratio $v_{bulk}/v_{random} = \tan\phi$. Dashed lines show light curves at $\phi = 30, 45$, and 60 degrees.

For comparison we also display a (standard) bulk motion magnification pattern in Figure 2b. Here a straight line at any angle corresponds to a microlensing lightcurve for bulk motion. Although the magnification pattern represents magnification as a function of position, once a straight track is chosen, the direction along this line can be interpreted as time ($t = x/v_{bulk}$). This means that one can choose a horizontal direction for the lightcurve, and thus directly compare the bulk motion pattern (Figure 2b) with the random motion pattern (Figure 2a): in both cases now the time direction is to the right. The two patterns look qualitatively different: for example, in the bulk motion pattern (Figure 2b) all the caustics (very dark lines, indicating formally infinite magnification) are concave, which is not true for the random motion pattern (Figure 2a).

## 3. Results

From the stack of 500 magnification patterns we can extract a large number of light curves for the two modes of motion: bulk and random. An example of each is shown on top of Figures 2a and 2b, where the horizontal line through the image indicates the position at which the lightcurve has been measured. The timescale of the lightcurve is in normalized time units $t/t_0$, where $t_0 = \xi_0/v$ is the time it takes to cross an Einstein radius. For easier comparison we assumed the same average velocity in both cases. Note that the peaks appear narrower in the random motion case. Here and for all subsequent light curves we assumed a Gaussian source profile with a width of $\sigma_S = 0.16\xi_0$ (or 4 pixels).

All the results shown here are based on a simulation with the surface mass density or optical depth in units of the critical surface mass density $\kappa = \Sigma/\Sigma_{cr} = 0.2$, where $\Sigma$ is the local physical surface mass density and $\Sigma_{cr} = \frac{c^2}{4\pi G}\frac{D_s}{D_d D_{ds}}$ is the critical surface mass density sufficient for multiple imaging. We decided to use a relatively low surface mass density in order to be able to distinguish and resolve individual microlensing peaks and to avoid crowding problems.

The main motivation of this paper is to find out how frequent or how important are high speed caustics or, equivalently, short duration events in the case of microlensing by random motion of stars. Our first paper on random motion (Kundić & Wambsganss 1993) concentrated on the direct analysis of magnification patterns. Here we analyze the microlensing light curves produced by sources with Gaussian profiles, which are more directly related to observations.

In multiply imaged quasars, gravitational microlensing is usually detected through peaks in the light curves of individual images; the maxima are characterized by high magnification (relatively large negative $\Delta m$) and short duration. The criteria to find an event depend on observational constraints, but the required magnification is typically at least $\Delta m \approx -0.2$ mag. The upper and lower limits on detectable event durations are set by the length of the monitoring program and the time resolution of the observations. Our peak finding algorithm imitates the observational situation as described above. Once the events have been identified, we can study the distribution of their widths (durations), peak magnifications (relative to shallower nearby minimum) and slopes.

In Figure 3a we display eight randomly chosen sample light curves (out of a total of 625) for random motion of lensing stars. For comparison we show a similar set of light curves for bulk motion in Figure 3b. The dotted horizontal lines mark the average magnification, while the vertical dotted lines indicate maxima identified by our peak–finding algorithm. It is quite obvious that the maxima for the random motion case are on average narrower, that their slopes are steeper, and that they are more frequent.

We also analyze the duration and the slope of the events for the two cases. We define the duration of an event as the full width at half maximum (FWHM), where the height of the half maximum is determined relative to the shallower of the two closest minima on either side of the maximum. Then we take the actual width of the peak at that value of the magnification. The slopes are defined as the absolute values of the changes in magnification between the maximum and the actual minima on either side, if they are steeper than a threshold value of 0.2 mag / 0.6 $\xi_0$. We use only one slope for a case with a steep increase on one side and a shallow decrease on the other. The determination of heights and slopes is illustrated in Figure 4.

The change in magnitude $|\Delta m|$, the "height" of a microlensing event, versus its FWHM is shown in Figure 5 for peaks found by our algorithm. Each point represents an identified peak with its width and height. There is a clear difference between the peaks of random motion (top) and bulk motion (bottom) light curves: the points in the top panel crowd much more around short durations. This is already evident in the sample light curves shown in Figures 3a and 3b. The reason for this is that the bulk motion light curves are limited by the size of the source, as mentioned above. This limitation does not apply for the random motion light curves, where the caustic velocity can be much larger than the projected transverse velocity. In fact, the short duration events originate from two different causes: when two lenses come close together, a pair of triangular caustics is created, which moves apart with a very high velocity. If a source is crossed by such a high-speed caustic, it can produce a peak that is very short. As the area of these triangles is quite small, however, the probability of being hit by such a caustic is not very high, and the magnification is not very large. Most of the short duration events come about from ordinary fold caustics that are moving (first expanding, then contracting) with a velocity a few times higher than the projected lens velocities, whenever two (or more) stars come close to each other in projection. Both of these effects can be easily seen in the accompanying video representation (cf. Appendix).

The excess of short events in the random motion light curves over the bulk motion light curves is evident in the plot of the distribution of event durations (Figure 6). The top panel displays the differential probability of the event durations. The solid line indicates the random motion light curves; the dotted line the bulk motion ones. The probability for the duration of random motion events is highest at about 0.2 $t_0$, whereas the probability distribution of bulk motion is much broader, peaking at about 0.5 $t_0$. The lower panel shows the corresponding cumulative distributions; here it can be seen that the median duration of the events are about 0.3 $t_0$ for the random motion and about 0.7 $t_0$ for the bulk motion. Figure 7 shows the probability distribution of the slopes of identified peaks. Again, random motion is indicated by solid lines, bulk motion by dotted lines. Top panel shows differential and bottom panel cumulative distributions. The difference between the curves in both Figures is quite obvious: for the case of random motion the peaks are clearly shorter and steeper.

As mentioned above, we expect realistic light curves to be produced by a combination of bulk motion and random motion. In general, random motions of galaxies are assumed to be larger than the velocity dispersions or the rotation velocities of stars in the disk. However, since for microlensing only the component transverse to the line of sight is important, it can occur that this component is quite small. The random velocity dispersion of stars in the lensing galaxy core, however, cannot become very small. Therefore, various combinations of the two velocity components are realized in nature. In Figure 8 we display three light curves for combinations of the two velocities: $\hat{v}_{bulk}/\hat{v}_{random} = 1/\sqrt{3}, 1, \sqrt{3}$ (corresponding to angles of $\phi = 30, 45, 60$ degrees), with fixed $\sqrt{\hat{v}_{bulk}^2 + \hat{v}_{random}^2} = v_{bulk} = v_{random}$. Even in these light curves it is suggestive that the steepness of the peaks increases with the contribution of random motion (i.e., from bottom to top).

We applied the same analysis to a simulation with the surface mass density of $\kappa = 0.5$ and confirmed our findings. The difference between random and bulk motion light curves seems to increase with $\kappa$: high surface density and random motion efficiently produce short, steep microlensing events.

## 4. Summary and Conclusions

We present and analyze microlensing light curves produced by random motion of individual stars in the lensing galaxy and compare the results with the "standard" approach to microlensing, assuming bulk motion of the galaxy only. This analysis is carried out for a moderate value of the surface mass density ($\kappa = 0.2$), because in this regime most microlensing events are still easy to identify.

After the paper was submitted for publication, higher surface mass density simulations ($\kappa = 0.5$ and $\kappa = 0.8$) were analyzed as well. The results qualitatively agree with the conclusions presented here and the figures are available via anonymous ftp to astro.princeton.edu (see

Appendix). With higher surface mass density, microlensing events last longer and have smaller amplitude due to the crowding of microlensing caustics. Frequently overlapping caustics make transitions from low amplification regions to high amplification regions more gradual. This is a very diferrent regime of microlensing than the low surface mass density case where most events are caused by single or double-star caustic crossing. At surface densities beyond critical ($kappa > 1$) we lack resolution to resolve individual events.

The main result of the paper is that microlensing by random motion of stars produces more events (almost twice as many), they are shorter, and the slopes are steeper than in the case of bulk motion. We understand the reasons for that qualitatively. In the case of bulk motion, the duration of an event cannot be shorter than the time it takes a fold caustic to cross the quasar. In this approach the relative velocity of quasar and caustic are given by the motion of the galaxy, projected onto the quasar plane. In the case of random motion of stars, however, the caustics change their shapes, whereby the velocities of fold caustics can be much larger than the projected velocity of the stars.

For observational purposes this means that one should think anew about strategies to observe microlensing. It may turn out to be more effective to monitor multiple quasars more frequently. This effect of microlensing by random motion of stars is always there, whereas the effect of microlensing by bulk motion of the galaxy can be small, in case the velocity component transverse to the line-of-sight is small.

We concentrate here on microlensing at optical depth of order unity. But in fact this effect of a changing lens configuration during the observation can be important for the searches for compact halo objects as well, since many of the lensing objects are double stars in the galactic disk or halo. Some of these binaries will be in the regime where the projected positions of the two stars change during the microlensing event. In this case it would not be possible to fit any static double lens lightcurve to the observed one.

We thank Bohdan Paczyński for encouragement in this project, for his reading of the manuscript and for his always helpful advice. We also thank Peter Schneider and Ed Turner for their comments on the manuscript. This project was supported by the NSF grant AST-9313620 and the STScI Grant #GO-2350-87A.

## 5. Appendix: Description of accompanying video

The accompanying video consists of four sequences: one header sequence, in which the film is described, and three sequences of 1000 magnification patterns calculated for random motion of stars, one sequence each for surface mass densities of $\kappa = 0.2, 0.5, 0.8$ (for the film we used twice as many frames as for the analysis – calculating an additional one between each two of the 500 frames analyzed above – in order to get smoother motion). The film shows the changing caustic network for stars moving with velocities drawn from a Maxwellian distribution, as described in the main part of the paper.

The video screen consists of four panels: the lower left shows the magnification pattern at any given time. The lower right displays positions of stars at that instant of time, so that one can correlate single, double, and multiple lens caustics with the corresponding lenses. It furthermore indicates velocity vectors of the lenses with small lines of different length and direction, so that one can see were the stars are heading. Both of these panels change with time. In addition, there are two panels at the top, indicating microlensing light curves. At the top left we show one lightcurve for bulk motion: it is just a horizontal cut through the magnification pattern (cf. Figure 1), i.e. the "standard" microlensing lightcurve as used in the past for fixed positions of the stars and only bulk motion of the lensing galaxy. Because the magnification pattern changes as time goes on, so does this lightcurve. This just means a different set of positions of the stars. An example for the new light curves determined for random motion of lenses is shown at the top right panel of the video screen. It is obtained at the central point of the magnification pattern. This lightcurve is fixed for the whole sequence; it is pre–calculated. All that changes is a black vertical line inside the lightcurve, indicating "time" or "frame number". In other words: for fixed positions of observer, lensing galaxy and quasar, the observer would see such a lightcurve with time due to the stars moving inside the galaxy.

After each such sequence for the three values of the surface mass density $\kappa$, the changing magnification pattern alone is shown in higher resolution, filling the whole screen (i.e., without the panels with positions of lensing stars and light curves). This is done in order to allow a more careful look at the way the caustics move and merge with each other.

Finally, the three sequences are shown again in a speed that is slowed down by about a factor of five, so that one can follow the evolution of individual caustics and the motion of the stars producing them. The total duration of the video is about 18 minutes. In Table 1 we indicate the parameters of the different sequences and their approximate durations.

The video qualitatively displays what is demonstrated in this paper: how the caustics move due to individual motion of stars, how they change, and how fast they change. To mention one example, it is quite interesting to see how the triangular caustics move apart with high speed, once two stars come close to each other in projection (cf. Schneider & Weiss 1986); in particular if there is some net angular momentum between these two stars (rather then head-on collision in projection). Then these caustics are "ejected" with a certain angular velocity; they leave the "main caustic" up to a maximum distance, keep rotating, and return to merge again with the leftover four-cusp caustic to build the six-cusp double lens caustic.

A short version of this video is available in digital format on the internet. In order to get access to the movie in MPEG format, please log in under anonymous ftp to astro.princeton.edu, give your userID as password, go to directory **jkw/microlensing/moving_stars**. There you will find additional figures from the analysis of $\kappa = 0.5$ and $\kappa = 0.8$ runs, as well as four parts of the MPEG movie: one each for the three values of $\kappa = 0.2, 0.5, 0.8$ plus one header sequence. These video clips can be looked at with the command "mpeg_play".

Table 1: **Parameters of Video Sequences:**

| Sequence | $\kappa$ | Contents | Duration |
|---|---|---|---|
| 0 | | header, description | 135 sec |
| 1a | 0.2 | magnification pattern, positions, 2 light curves | 60 sec |
| 1b | 0.2 | magnification pattern (high resolution) | 60 sec |
| 2a | 0.5 | magnification pattern, positions, 2 light curves | 60 sec |
| 2b | 0.5 | magnification pattern (high resolution) | 60 sec |
| 3a | 0.8 | magnification pattern, positions, 2 light curves | 30 sec |
| 3b | 0.8 | magnification pattern (high resolution) | 30 sec |
| 1a | 0.2 | magnification pattern, positions, 2 light curves | 220 sec |
| 2a | 0.5 | magnification pattern, positions, 2 light curves | 220 sec |
| 3a | 0.8 | magnification pattern, positions, 2 light curves | 220 sec |

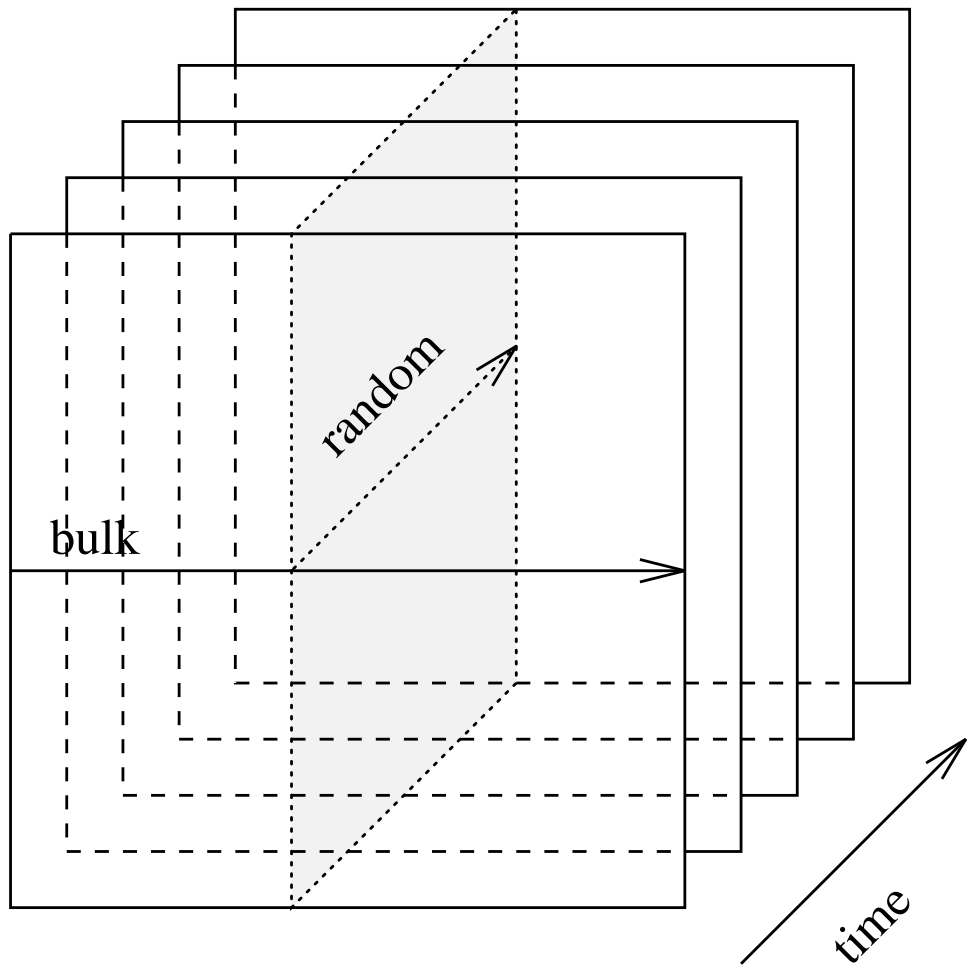

Fig. 1.— This stack of squares symbolizes our method of calculating light curves for microlensing by random motion of stars. Each square represents a magnification pattern for a fixed stellar field. Adjacent squares contain magnification patterns for slightly different star positions. Displayed are five squares; in fact we have calculated 500 slightly different magnification patterns. Light curves for microlensing by bulk motion are obtained by any straight line *inside* one square. As an example we show a horizontal line with an arrow, labelled "bulk". Light curves for pure random motion are obtained by fixing a position at one square, and then moving through the whole stack of squares at always the same position. This is indicated by the dotted line with an arrow, labelled "random". The shaded area (square in perspective) represents a whole set of light curves for random motion, for lines parallel to the dotted line.

This figure can be obtained via anonymous ftp from astro.princeton.edu,
cd jkw/microlensing/moving_stars
get fig2.ps.Z

Fig. 2.— An example of the microlensing pattern used to extract random motion light curves. The gray intensity indicates the magnification at the particular location (darker means higher magnification). The field shown in this figure corresponds to the shaded region in Figure 1. For microlensing by pure random motion, one can interpret this figure as follows: time corresponds to the horizontal direction and space to the vertical direction. The lightcurve shown on top is for an extended source with a Gaussian profile of width $\sigma = 0.16 \xi_0$, and it is extracted along the horizontal line marked in the figure. Light curves along parallel lines would represent light curves for pure random motion, but different source positions. The dashed lines indicate light curves for different combinations of random and bulk motion, as displayed in Figure 8.

This figure can be obtained via anonymous ftp from astro.princeton.edu:
cd jkw/microlensing/moving_stars
get fig3.ps.Z

Fig. 3.— For comparison with Figure 2a we show here a standard microlensing pattern for a *fixed* stellar configuration, i.e. for bulk motion of the lensing galaxy. The lightcurve at the top is extracted along the horizontal line marked in the figure (same source size as above). Equivalently, the horizontal dimension represents time in this case. Note the qualitative differences in the structure of the high magnification regions here and in Figure 2a.

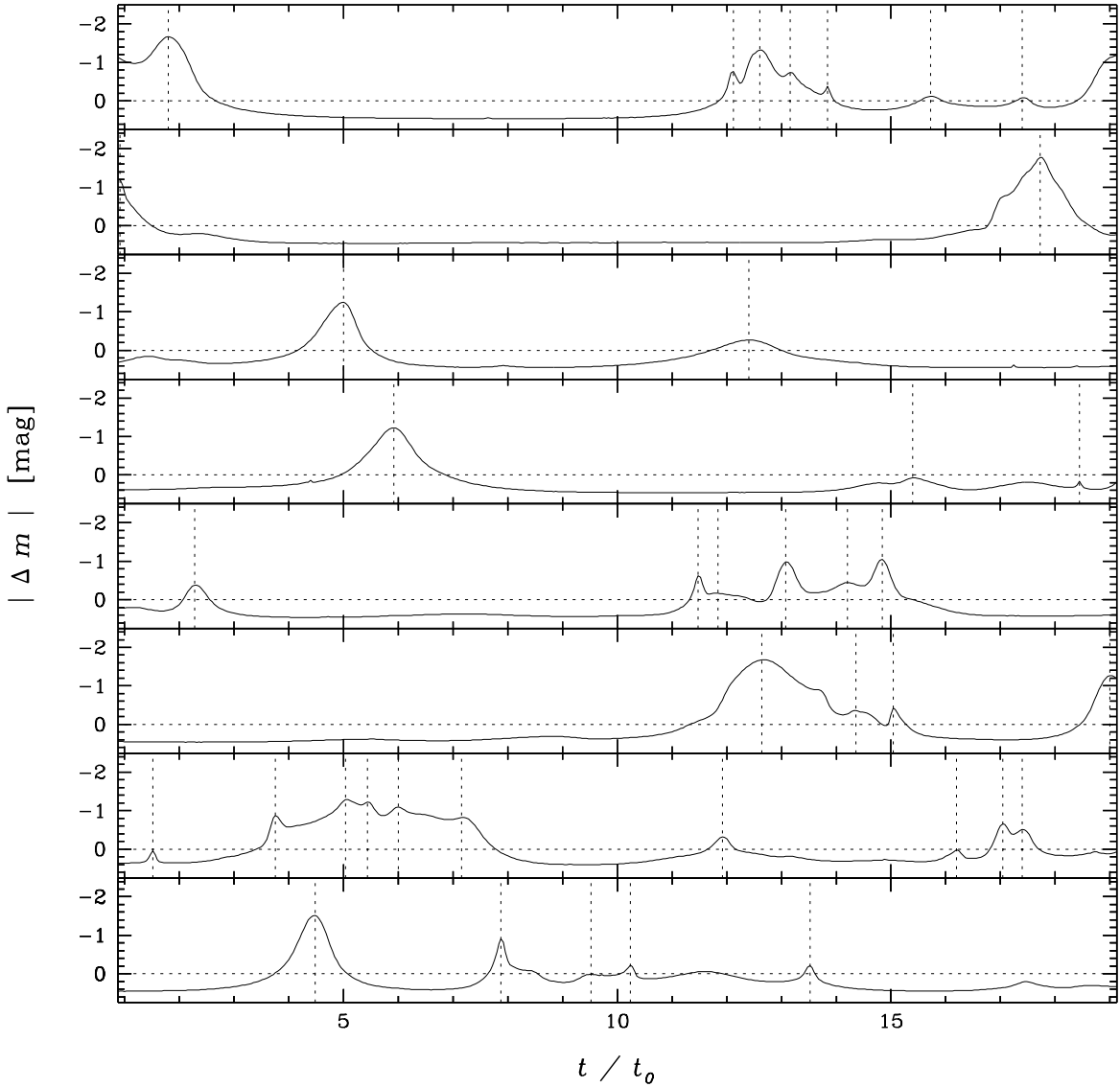

Fig. 4.— Eight arbitrarily chosen light curves corresponding to microlensing by random motion. The time scale is in normalized time units $t/t_0$, where $t_0 = \xi_0/v$ is the time it takes to cross one Einstein radius. The dotted horizontal lines indicate average magnification, the dotted vertical lines show microlensing events that were identified with our peak-finding algorithm. Note that microlensing peaks here are more frequent and steeper than those of bulk motion (Figure 3b).

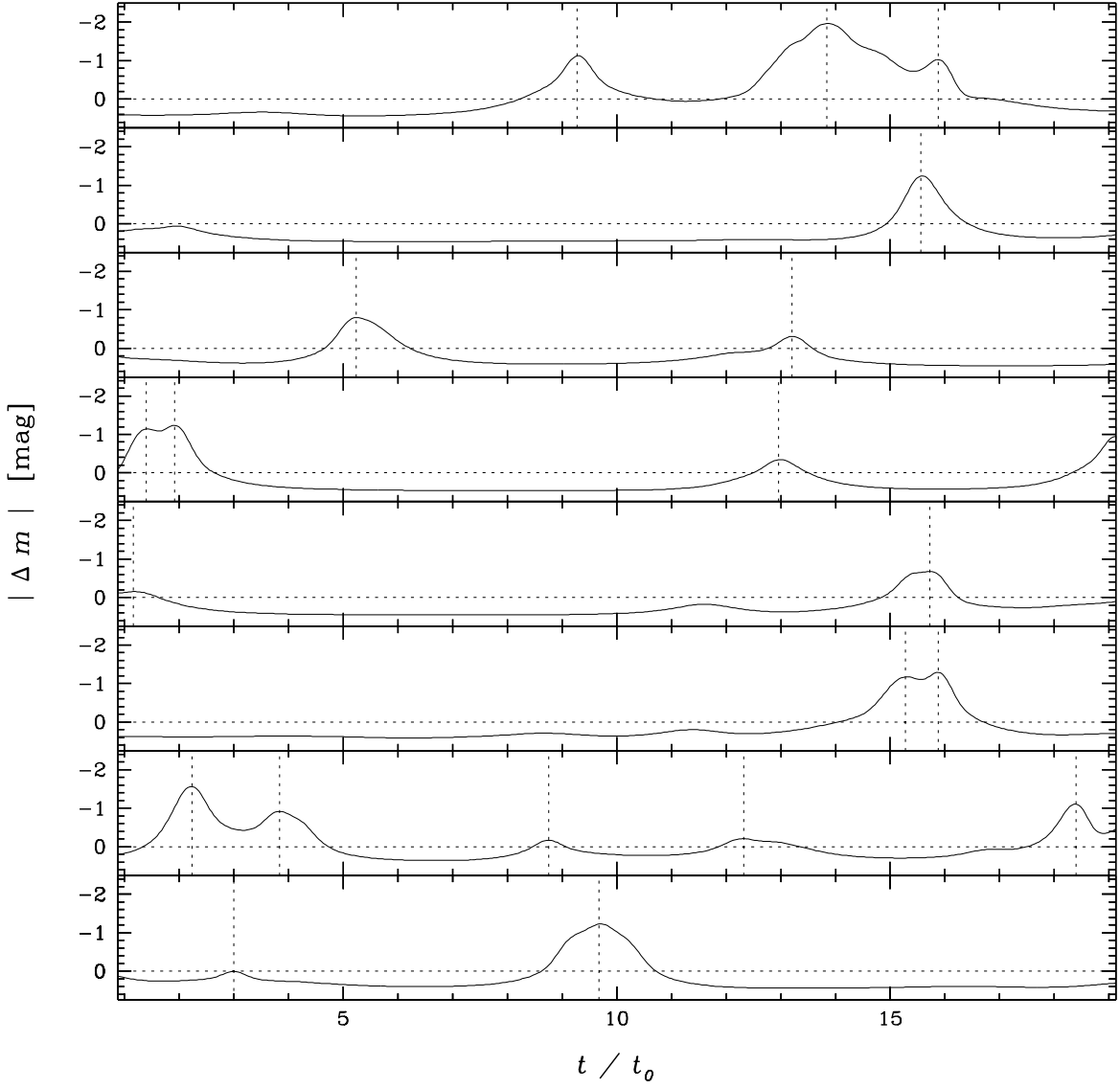

Fig. 5.— Same as Figure 3a for microlensing by bulk motion of the lensing galaxy.

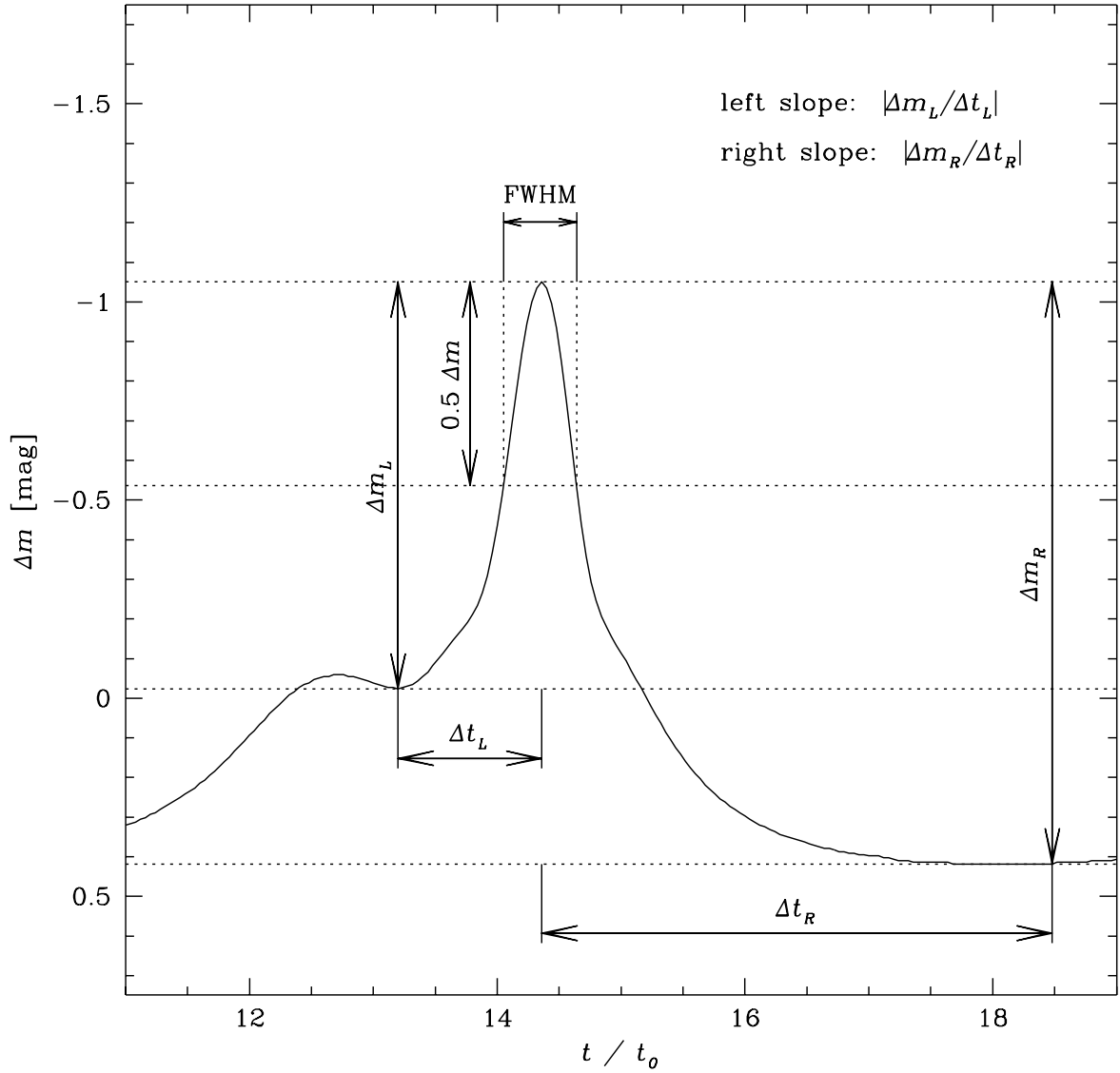

Fig. 6.— A sample microlensing event identified by our peak–finding algorithm. Magnitude differences $\Delta m$ and durations $\Delta t$ are plotted on both sides of the asymmetric event. The higher (left) baseline is used to define half maximum and full width at half maximum (FWHM).

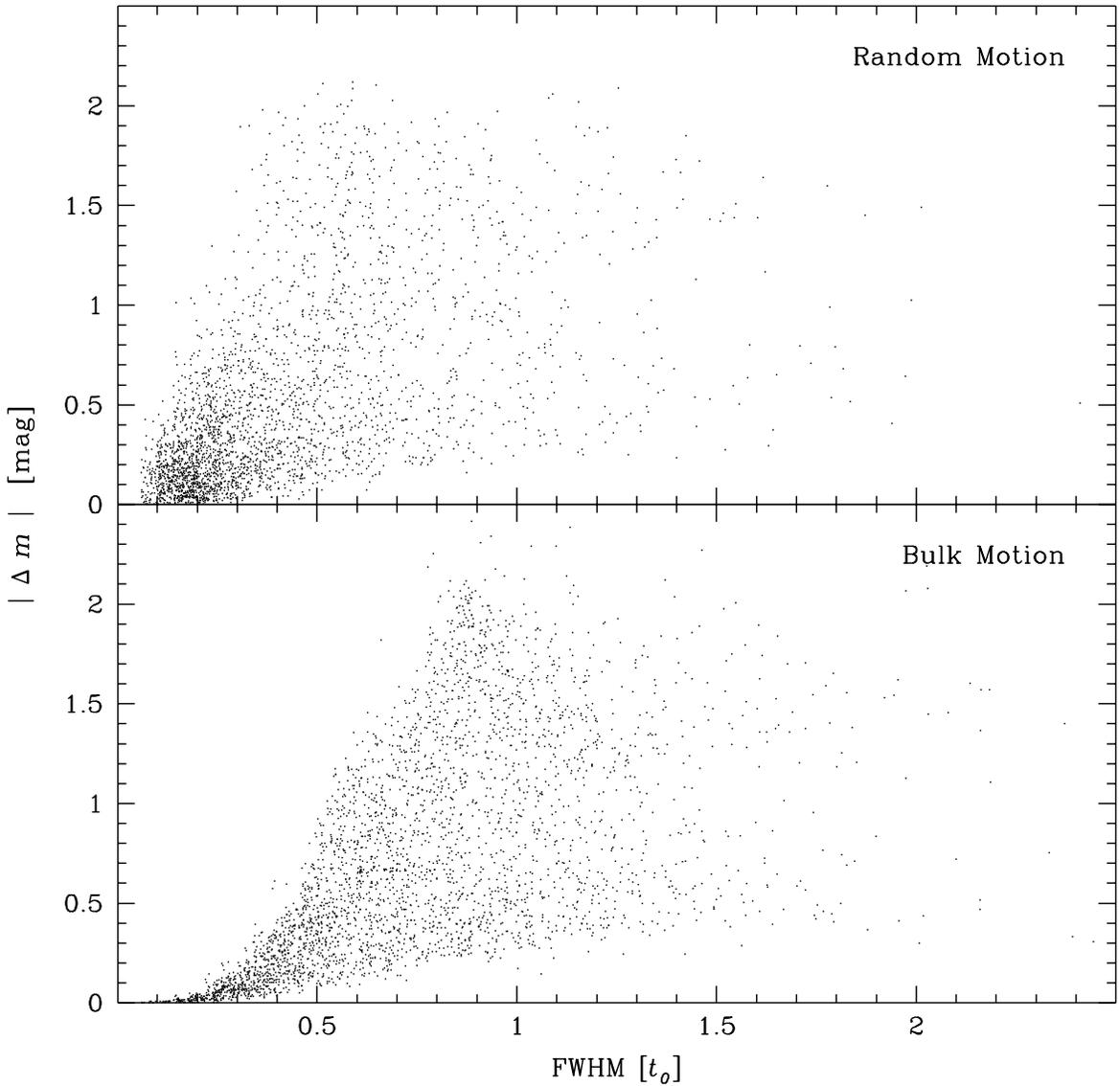

Fig. 7.— Distribution of heights and durations (full width half maxima) of identified microlensing events (top panel: random motion, bottom panel: bulk motion). Height is defined as twice the half maximum.

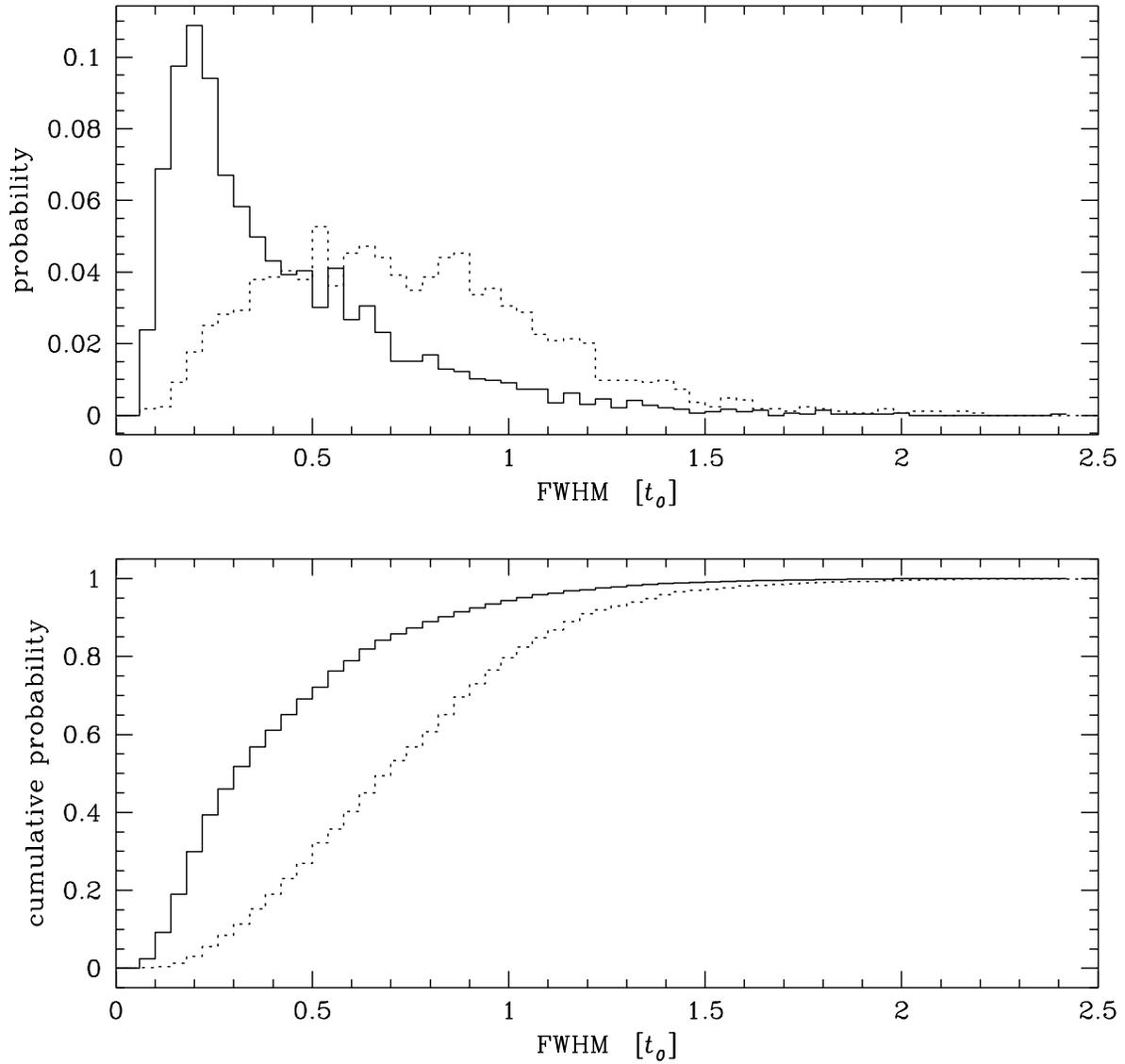

Fig. 8.— Differential (top) and cumulative (bottom) statistics of microlensing event *durations* defined as FWHM (see text). The solid line is for random motion events, the dotted line is for bulk motion events.

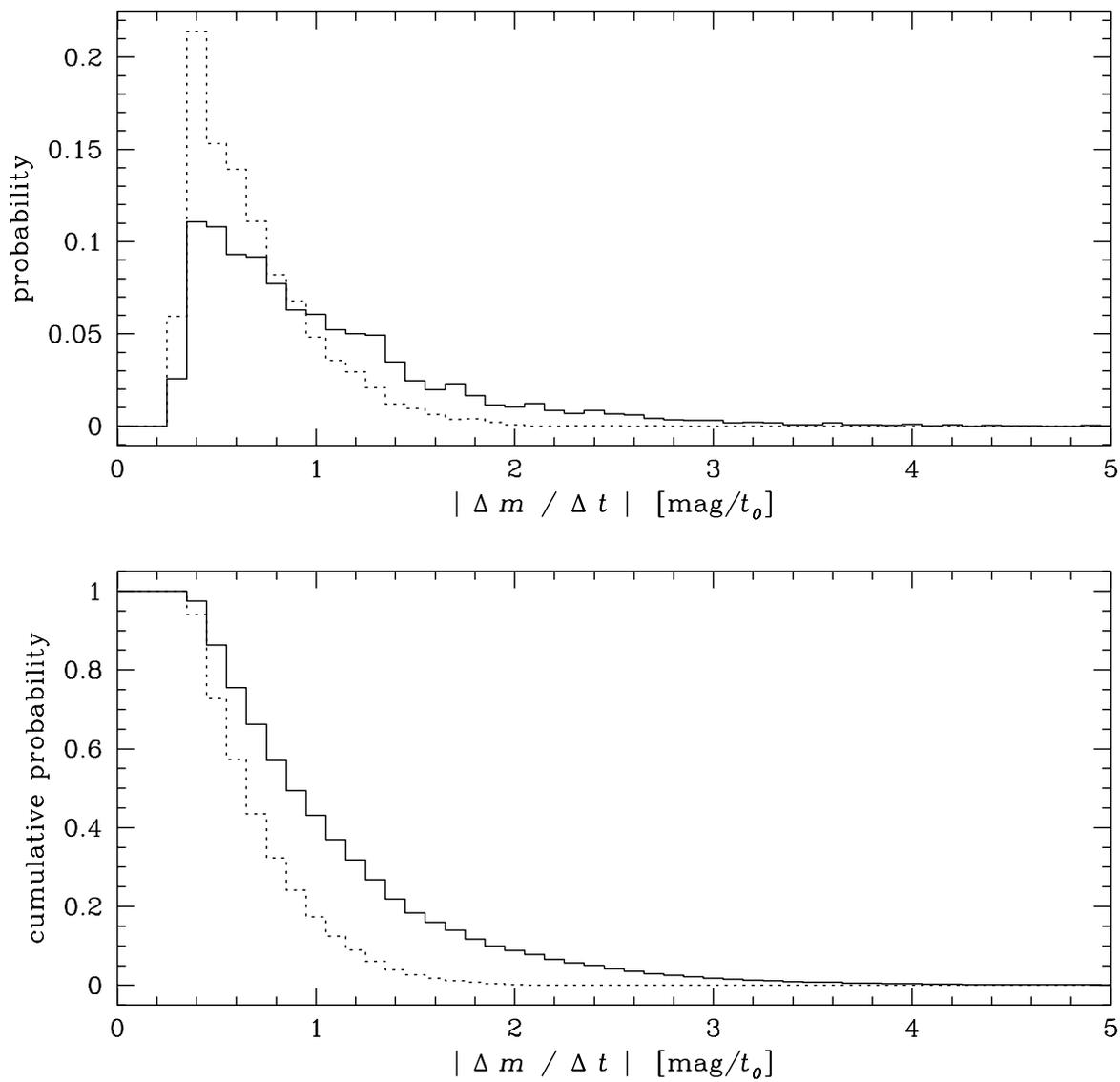

Fig. 9.— Differential (top) and cumulative (bottom) statistics of microlensing event *slopes*. The solid line is for random motion events; the dotted line is for bulk motion events.

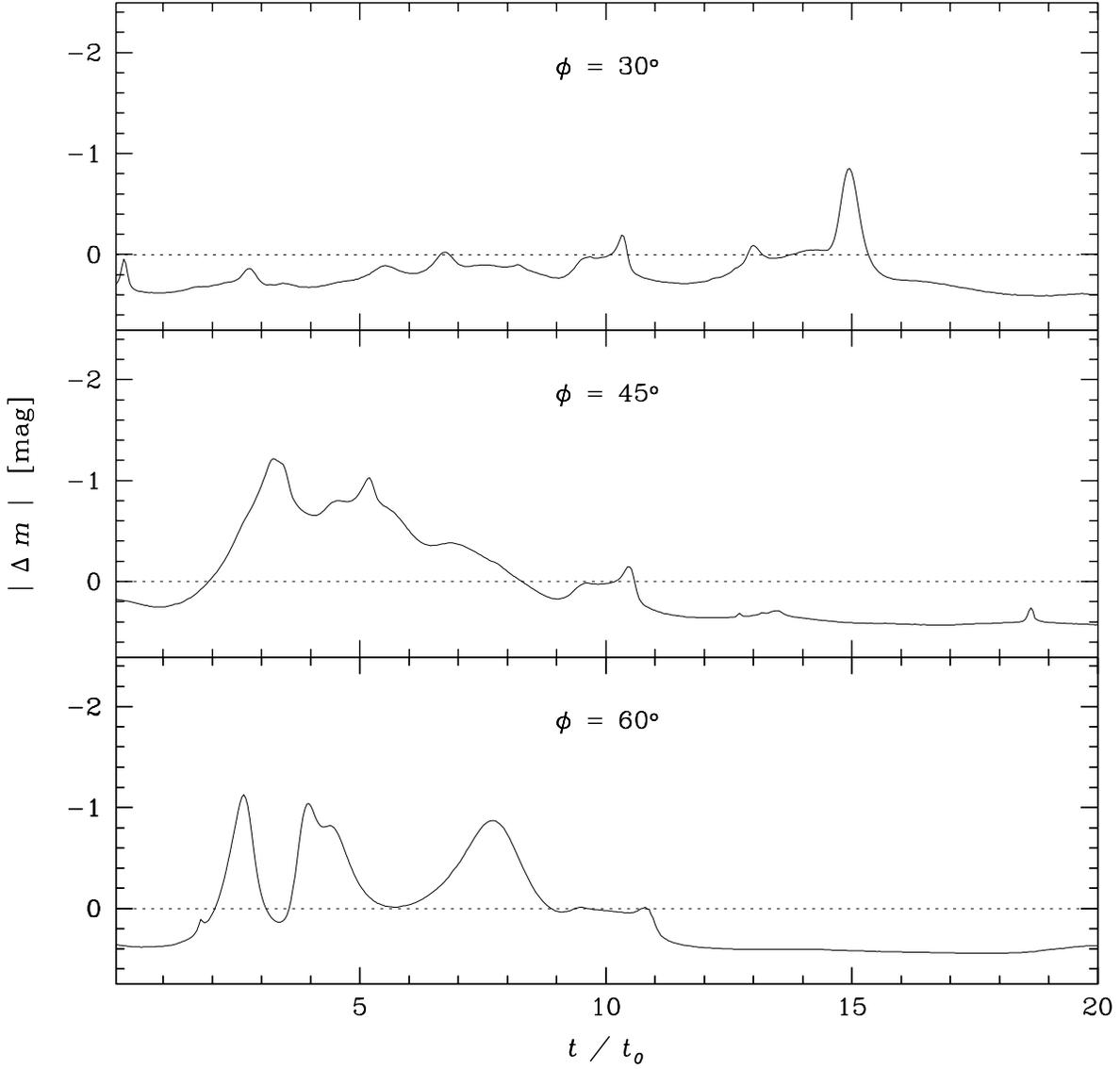

Fig. 10.— Light curves for certain combinations of random and bulk motion: $v_{bulk}/v_{random} = 1/\sqrt{3}, 1, \sqrt{3}$ (corresponding to angles of $\phi = 30, 45, 60$ degrees). The tracks along which these light curves are determined are indicated with dashed lines on Figure 2b. Even in these light curves it is suggestive that the steepness of peaks increases with the contribution of random motion (i.e., from bottom to top).